\documentclass[12pt]{article}
\usepackage{epsfig,latexsym}

\def\be{\begin{equation}}
\def\ee{\end{equation}}

\def\appendix#1{\addtocounter{section}{1}\setcounter{equation}{0}
\renewcommand{\thesection}{\Alph{section}}
\section*{\protect\noindent \parbox[t]{16cm}{#1}}
\addcontentsline{toc}{section}{ \thesection\ \ \ #1}}

\def\mun{{\underline{m}}}

\def\cI{{\cal I}}
\def\cJ{{\cal J}}

\def\x'{\mathaccent 19 x}
\def\th'{\mathaccent 19 \theta}
\def\barth'{\mathaccent 19 {\bar{\theta}}}

\def\del{\partial}

\def \a {\alpha}

\def \s {\sigma}

\def \G {\Gamma}

\def \g {\gamma}

\def \e#1 {{\rm e}^{#1}}

\def \ha { { 1\over 2 }}

\def \ov {\over}

\def\Z'{\mathaccent 19 Z}

\def \g {\gamma}
\def \G {\Gamma}

\def \la {\label}

\def \lc { light-cone }

\def\cI{{\cal I}}

\def\cJ{{\cal J}}

\def \mm {{\rm m}}

\def \f {{\rm f}}

\def\mun{{\underline{m}}}

\def \lc {light-cone\ }

\def \s { \sigma }

 \def \a { \alpha}

\def \lc {light-cone\ }

\begin{document}
\font\cmss=cmss10 \font\cmsss=cmss10 at 7pt
\hfill IC/2002/36 \vskip .1in \hfill Bicocca-FT-02-8 \vskip .1in \hfill hep-th/0205296

\hfill
\vspace{18pt}
\begin{center}
{\Large \textbf{PP-wave and Non-supersymmetric Gauge Theory}}
\end{center}

\vspace{6pt}

\begin{center}
{\textsl{F. Bigazzi $^{a}$, A. L. Cotrone $^{b}$, L. Girardello $^{b}$ and A. Zaffaroni $^{b}$}}

\vspace{20pt}

\textit{$^a$ The Abdus Salam ICTP, Strada
Costiera, 11; I-34014 Trieste, Italy.}

\textit{$^b$ Dipartimento di Fisica, Universit\`{a} di Milano-Bicocca, P.zza della Scienza, 3; I-20126 Milano, Italy.}

\end{center}

\vspace{12pt}

\begin{center}
\textbf{Abstract }
\end{center}

\vspace{4pt} {\small \noindent We extend the pp-wave correspondence to a non 
supersymmetric example. The model is the type 0B string theory on 
the pp-wave R--R background. We explicitly solve the model and 
give the spectrum of physical states. 
The field theory counterpart is given by a sector of the large N SU(N) 
$\times$ SU(N) CFT living on a stack of N electric and N magnetic D3-branes. 
The relevant  effective coupling constant is $g_{eff}=g_sN/J^2$. 
The string theory has a tachyon in the spectrum, whose light-cone energy
can be exactly computed as a function of $g_{eff}$.
We argue that the perturbative analysis in $g_{eff}$ in the dual gauge theory
is reliable, with corrections of non perturbative type. 
We find a precise state/operator map, showing that the first perturbative corrections to the anomalous dimensions of the operators have the behavior expected from the string analysis.}
\vfill
\vskip 5.mm
 \hrule width 5.cm
\vskip 2.mm
{\small
\noindent
bigazzif@ictp.trieste.it\\
aldo.cotrone@mib.infn.it\\
luciano.girardello@mib.infn.it\\
alberto.zaffaroni@mib.infn.it}
\section{Introduction and conclusions}

In \cite{MBN} Berenstein, Maldacena and Nastase (BMN) studied the
correspondence between type IIB string theory and the large N, 
${\cal N}=4$ SU(N) SCFT beyond the supergravity approximation. 
They found a precise map between {\it all the string modes} on a 
maximally supersymmetric plane wave background, and a sector of
gauge invariant operators of the field theory.
The string states are the ones with very large angular momentum 
($ J\sim \sqrt N$) along the equator of $S^5$ in the original 
$AdS_5 \times S^5$ background, which reduces to the pp-wave in the Penrose limit \cite{Blau}. The resulting spacetime, which is supported by a null, constant five from field strength, has the form:
\begin{eqnarray}
ds^2 &=& 2dx^+dx^- - f^2x^2dx^+dx^+ + dx^Idx^I , \quad I=1,..,8 \nonumber \\
&&F_{+1234}= F_{+5678}= 2f .
\label{pp}
\end{eqnarray}
The symmetry of the solution is 
$SO(4)\times SO(4)$.

Despite the presence of the R--R field, type IIB string theory is exactly solvable on the pp-wave \cite{M}. The 
spectrum, in light cone gauge, is obtained by acting on the vacuum with
free oscillators.  The light-cone
Hamiltonian is 
\be \label{prima}
H=p_+ = \sum_{n=-\infty}^\infty N_n \sqrt {f^2 + {n^2\over
(\alpha' p^+)^2 } }
\ ,
\ee
where $N_n$ is the occupation number of the $n$-th oscillator.
This formula implies 
the following relation between the dimension $\Delta$ and the
R-charge $J$ of the corresponding gauge theory operator:
\be \label{seconda}
\Delta - J =\sum_{n=-\infty}^\infty N_n \sqrt {1 + {4\pi g_s N n^2\over
J^2} }
\ .
\ee
This expression, valid for $\frac{\Delta-J}{J} \ll 1$, follows from the
relations $p_+ = f(\Delta - J)$ and $p_- \approx {J\over fR^{2}}$, 
where $R^4\equiv 4\pi{\alpha'}^2g_sN$ is the common radius of $AdS_5$ 
and $S^5$.
In the Penrose limit $R$ (and so $g_sN$), together with $J$ and $\Delta$, go to infinity, in such a way that $p_+$ and $p_-$ (and so $\Delta - J$ and ${J^2\over g_sN}$) stay finite.

BMN were able to identify the single-trace field theory operators
with scaling dimensions given by (\ref{seconda}). 
The operators corresponding to
zero-mode states are chiral, so their scaling dimension is the
same as in free field theory. The remaining operators are
non-chiral, but their anomalous dimensions are predicted to be finite even 
in the strong coupling limit.
The perturbative expansion for this class of
almost BPS states is in fact controlled by $g_{eff}={\lambda \over J^2}=
{1\over 4\pi(f\alpha ' p^+)^2}$ rather than by the (large)
't Hooft coupling $\lambda=g_sN$. Anomalous dimensions can be 
perturbatively evaluated by expanding around the free field theory
with effective coupling $g_{eff}$. 
\\

In this paper we analyze the pp-wave correspondence for a non supersymmetric 
model. The natural candidate to study is the SU(N) $\times$ SU(N) gauge theory introduced in \cite{KT}. It is dual to type 0B string theory on $AdS_5 \times S^5$, obtained as the near horizon geometry of a stack of N electric and N magnetic D3-branes. The gravity background supports a self-dual 5-form field strength, a constant dilaton and a constant (namely zero) tachyon field. We can thus perform the Penrose limit exactly as in the case of the type IIB theory, ending up with type 0B on the pp-wave background (\ref{pp}).
The theory consists of an untwisted sector, which is exactly the same as the 
bosonic part of the parent type IIB string, and a twisted one, which has a 
tachyonic ground state.
The light cone energy of the tachyon is a function of $g_{eff}$. It
is a negative definite function, it diverges 
in the limit $g_{eff} \rightarrow \infty$,
which closely mimics the flat space case,
and it vanishes in the opposite free field theory limit 
$g_{eff}\rightarrow 0$. 

This result deserves some comments. The original theory 
before the Penrose limit,
type 0B on $AdS_5 \times S^5$, has a tachyonic 
instability at large $\lambda$ associated with a complex dimension operator 
in the dual field theory \cite{K}. 
However, this operator could be removed from the spectrum in the Penrose 
limit, as happens, for example, 
in \cite{mukhi}. The Penrose limit indeed  selects only the operators 
with large, positive mass
and scaling dimension $\Delta \sim \sqrt{\lambda}$.
We can expect a similar behavior in our case, since the large 
and positive contribution of the angular momentum ($\sim J^2=O(\lambda)$) should overwhelm the negative mass of the tachyonic ground state ($\sim -\sqrt{\lambda}$). 
Thus, all complex dimension field theory operators are not expected to survive the Penrose limit and 
$H_{AdS}\sim\Delta$ should be well-defined.
The original instability of the $AdS$ background is reflected in the fact 
that the spectrum of the pp-wave exhibits a 
light-cone  energy $H_{pp-wave}\sim \Delta-J$ 
which can be infinitely negative 
in the gravity limit. 
It would be very interesting to understand if this unboundness of the 
light cone energy corresponds to a real instability in the string theory and
to study its nature and origin directly in $AdS$ and in the 
field theory.

One of the interesting aspects of the pp-wave background is that 
the energy of the tachyon as a function of 
the coupling $g_{eff}$ can be exactly computed. 
As already mentioned, the light cone energy 
of the tachyon is always negative.  
More interestingly, the energy is zero for $g_{eff}=0$
and it is perturbatively zero to all orders in $g_{eff}$.
This means that the corresponding field theory is perturbatively stable around 
the free field theory limit.
We conclude that we can study the correspondence 
between strings and operators in a $non-supersymmetric$ theory 
order by order in a Feynman diagram expansion, as was done in the 
supersymmetric case in \cite{MBN}.
We can compare the behavior of the tachyon energy 
with the analogous situation in $AdS$.
The mass of the tachyon in type 0B on $AdS$ is a 
function of $\lambda$ and, as was 
argued in \cite{KT}, it should become positive for 
$\lambda$ small enough, i.e. 
in these theories one can expect a transition 
from a stable to an unstable regime by varying the coupling constant.
The transition is due to the effect of the RR five-form field, which
contributes to the effective mass of the tachyon.
In the pp-wave background 
there is no contribution from the null five-form field 
and the energy is always negative.
\\

The paper is organized as follows. After a brief review of the type IIB GS 
formalism on the pp-wave in Section two, in Section three we study the type 
0B string on the same background. The theory is obtained as an orbifold of 
type IIB by $(-1)^F$, $F$ being the spacetime fermion number. It has an 
untwisted and a twisted sector. In both of them we calculate the light-cone 
Hamiltonian and we find that in the twisted sector the lowest state has 
negative energy. In Section four the spectrum of physical states is presented.
In Section five we identify the gauge theory operators dual to the string 
states. 
We compare the predictions of the string Hamiltonian with
the planar corrections to the dimensions of the operators for small $g_{eff}$,
as suggested in \cite{MBN}, finding agreement both in the untwisted and
twisted sector.
We are not able to reproduce the part of the corrections to the classical 
dimension of the operators corresponding to the energy of the tachyon. 
It appears to involve a non-perturbative calculation in $g_{eff}$. 
It would be interesting to investigate the twisted  sector at nonzero string coupling \cite{altri,gross}, in order to understand better
the issue of stability in these type of non supersymmetric models. 
Finally, the identification of the ground state with a tachyon 
is reproduced by a 
gravity calculation in the Appendix.

\section{Type IIB GS superstring on pp-wave}
In this paper we follow the conventions of \cite{M,MT}.
The Green-Schwarz action for type IIB superstring on the
background (\ref{pp}) has a very simple form in light-cone gauge
\cite{M}, where the following conditions (in units $2\pi\alpha'=1$) are imposed:
\be \Gamma^+\theta^1=
\Gamma^+\theta^2=0,\quad    x^+ = p^+\tau . \label{lcg} \ee Here
$\theta^{1a},\theta^{2a}, a=1,...,16$ are two ten dimensional
Majorana-Weyl spinors of the same chirality, and: \be \Gamma^{\pm}=
{1\over\sqrt2}(\Gamma^9 \pm \Gamma^0), \quad (\Gamma^{\pm})^2=0,
\ee implying that
only a half of the components of the fermionic fields are dynamical
variables in \lc gauge. Among the bosonic degrees of freedom, only the eight
transverse ones, $x^I, I=1,..,8$, are independent.
The gauge-fixed equations of motion for the physical variables can be derived from the action:
\be
S=\int d\tau d\sigma\left[ \ha(  \del_+ x^I \del_- x^I  -  \mm^2  x^2_I)+{\rm i}  ( \theta^1\bar{\gamma}^- \partial_+  \theta^1   +
\theta^2 \bar{\gamma}^-\partial_-  \theta^2 -2 \mm    \theta^1
\bar{\gamma}^- \Pi \theta^2 )\right],
\label{action}
\ee
(where $\partial_\pm\equiv \partial_{\tau} \pm \partial_{\sigma}$, $\mm \equiv p^+ \f$ and $\bar \gamma^+\theta^{\cI}=0$\footnote{ Here, just as in \cite{M,MT}, $ \g^m , \
\bar \g^m  $ are  the $16 \times 16$ Dirac matrices
which are the off-diagonal parts of $32 \times 32 $   matrices  $\G^m$.
The matrix  $\Pi$ ($\Pi^2 =1$) is the product of four $\gamma$-matrices and have its origin from the coupling with
the five form background field strength.
}) supported by the constraint:
 \be \int d\sigma
[\partial_{\tau}x^I\partial_{\sigma}x^I + {\rm
i}(\theta^{1}\bar{\gamma}^-\partial_{\sigma}\theta^{1}+
\theta^{2}\bar{\gamma}^-\partial_{\sigma}\theta^{2})]= 0,
\label{zeromoml} \ee
which will enforce invariance under $\sigma$ translations.
The action (\ref{action}) reduces to the light-cone GS one in flat spacetime when $\f=0$. When $f\ne0$ it simply describes eight free massive 2d
scalar fields $x^I$  and eight free massive 2d Majorana
fermionic fields (with real components $\theta^1,\theta^2$) whose
equations of motion read: \be \del_+ \del_- x^I  + \mm^2 x^I=0\,
,\qquad
\partial_+ \theta^1  -
\mm \Pi \theta^2 =0\,,\qquad \partial_- \theta^2  +
\mm \Pi \theta^1 =0\,. \label{equamot}
\ee
With the use of the fermionic equations of motion, the light-cone Hamiltonian can be cast in the form \cite{M}:
\be
H ={ 1 \ov p^+} \int_0^1 {d\s}
 \ \big[ \frac{1}{2 } (\pi_I^2 + (\partial_{\sigma}x_I)^2+
\mm^2 x_I^2) +{\rm i}(\theta^1 \bar{\gamma}^- \partial_{\tau}\theta^1
+\theta^2\bar{\gamma}^-\partial_{\tau}\theta^2) \big] \ ,
\label{ham2}
\ee
where $\pi_I=\partial_{\tau}x_I$ are the bosonic canonical momenta.

\section{Type 0B model and its quantization}
In order to obtain the type 0B model we perform a
quotient of the type IIB theory by  $(-1)^F$, where $F$ is
the spacetime fermion number. In the GS formulation this
corresponds to the projection:
\begin{equation}
\theta \approx -\theta
\ee
on the fermionic coordinates. Just as in ordinary orbifolds, this gives us two sectors.

In the {\bf untwisted sector} worldsheet bosons and fermions obey
periodic boundary conditions: \be x^I(\sigma+1,\tau) =
x^I(\sigma,\tau)\,,\qquad
\theta(\sigma+1,\tau)=\theta(\sigma,\tau)\,,\qquad \qquad 0\leq
\sigma \leq 1\ . \ee From these one can get \cite{MT} the
following general solution of the equations of motion
(\ref{equamot}): \be\label{xIsol} x^I(\sigma,\tau) = \cos \mm
\tau\, x_0^I +\mm^{-1} \sin \mm \tau\, p_0^I + {\rm
i}\sum_{n=\!\!\!\!/\,\, 0 }\frac{1}{\omega_n}\Bigl(
\varphi_n^1(\sigma,\tau) \alpha_n^{1I} + \varphi_n^2(\sigma,\tau)
\alpha_n^{2I}\Bigr), \ee \be \label{th1sol}\theta^1(\sigma,\tau)
=\cos\mm\tau\, \theta_0^1 +\sin\mm\tau\  \Pi\theta_0^2+
\sum_{n=\!\!\!\!/\,\, 0}c_n\Bigl( \varphi_n^1(\sigma,\tau)
\theta^1_n + {\rm i}{\textstyle{\omega_n-k_n\ov
\mm}}\varphi_n^2(\sigma,\tau)\Pi \theta_n^2\Bigr), \ee
\be\label{th2sol} \theta^2(\sigma,\tau) =\cos\mm\tau\, \theta_0^2
- \sin\mm\tau\ \Pi\theta_0^1+ \sum_{n=\!\!\!\!/\,\, 0} c_n\Bigl(
\varphi_n^2(\sigma,\tau) \theta^2_n - {\rm
i}{\textstyle{\omega_n-k_n\ov \mm}}\varphi_n^1(\sigma,\tau)\Pi
\theta_n^1\Bigr), \ee where: \be\varphi^1_n(\sigma,\tau) =
\exp(-{\rm i}(\omega_n \tau -k_n\sigma))\,,\qquad
\varphi^2_n(\sigma,\tau) = \exp(-{\rm i}(\omega_n \tau +
k_n\sigma))\ee and

\be \omega_n =\sqrt{k_n^2 + \mm^2 },\quad n>0\ ;\qquad
 \omega_n
=-\sqrt{k_n^2 + \mm^2 },\quad  n < 0\ ;\ee \be \label{cn} k_n
\equiv 2\pi n \ , \ \ \ \ \ \ \ \ \ \ \ \  \  c_n
=\frac{1}{\sqrt{1+({\omega_n -k_n\ov \mm})^2}}\,, \qquad n=\pm
1,\pm 2,\ldots \ .  \ee

In the {\bf twisted sector} the worldsheet fermions get antiperiodic boundary conditions:
\be
\theta(\sigma+1,\tau)= - \theta(\sigma,\tau)\,,\qquad \qquad 0\leq
\sigma \leq 1\ . \ee
Thus the corresponding solutions of the equations of motion are (those for the bosons are exactly the same as in the untwisted sector):
\be \label{Tth1sol}\theta^1(\sigma,\tau) = \sum_{r}c_r\Bigl(
\varphi_r^1(\sigma,\tau) \theta^1_r + {\rm
i}{\textstyle{\omega_r-k_r\ov \mm}}\varphi_r^2(\sigma,\tau)\Pi \theta_r^2\Bigr), \ee
\be\label{Tth2sol} \theta^2(\sigma,\tau) = \sum_{r} c_r\Bigl(
\varphi_r^2(\sigma,\tau) \theta^2_r - {\rm
i}{\textstyle{\omega_r-k_r\ov \mm}}\varphi_r^1(\sigma,\tau)\Pi \theta_r^1\Bigr). \ee
Here the index $r$ runs over the {\it half-integer} numbers, and, due to the antiperiodic boundary condition, no zero mode appears.

The quantization of this model is straightforward. The physical
states are required to be invariant under $\theta\rightarrow -\theta$
and to have zero-eigenvalue under the operator (\ref{zeromoml}). 

The Hamiltonian (\ref{ham2}) aquires a very simple form in terms of the following operators (see \cite{MT}):

\be a_0^I = \frac{1}{\sqrt{2\mm}}(p_0^I + {\rm
i}\mm x_0^I)\,,\qquad \bar{a}_0^I =\frac{1}{\sqrt{2\mm}}( p_0^I -
{\rm i}\mm x_0^I)\,, \label{bososc} \ee

\be \alpha_{-n}^I =\sqrt{\frac{\omega_n}{2}}\,a_n^I\,,\qquad
 \alpha_n^I =\sqrt{ \frac{\omega_n}{2}}\,\bar{a}_n^I\,,\qquad n =1,2,\ldots \label{nbososc} \ee

\be \label{nss} \theta_R =   \frac{1 +\Pi}{2} (\theta^1_0 + i
\theta^2_0)
\ , \ \ \ \ \ \ \ \ \ \theta_L =  \frac{1-\Pi}{2}(\theta^1_0 + i
\theta^2_0)
 \ , \ee
\be \label{ddo}
\theta^\cI_{-q}\equiv \frac{1}{\sqrt{2}}\eta^\cI_q  \ ,
\qquad \theta^\cI_q\equiv
\frac{1}{\sqrt{2}}\bar{\eta}^\cI_q\, , \ee
(here q is integer and not zero, or half-integer) whose (anti)commutation algebra reads:
\be
[\bar{a}_0^I, a_0^J]=\delta^{IJ} \ , \ \ \ \ \ \ \
 [\bar{a}_m^{\cI I},a_n^{\cJ
J}]=\delta_{m,n}\delta^{IJ}\delta^{\cI\cJ}\,,\ee \be \la{pip}
\{\bar{\eta}_q^{\cI}{}^a,\eta_p^{\cJ}{}^b\}
=\frac{1}{2}(\gamma^+)^{ab}\delta_{q,p}\delta^{\cI\cJ}\,,\qquad
\{\bar{\theta}_R,\theta_L\}=0 \,,\ee \be \la{pipo}
\{\bar{\theta}_R,\theta_R\}=\frac{(1+\Pi)}{4}\gamma^+ \,,\qquad
\{\bar{\theta}_L,\theta_L\}=\frac{(1-\Pi)}{4}\gamma^+
 \ . \ee

Correspondingly we define the Fock vacuum in each sector as:
\be\label{vacdef1} \bar{a}_0^I|0, p^+\rangle_{U,T}=0\,,\qquad \bar{a}_n^{\cI
I}|0, p^+\rangle_{U,T}=0\,,\qquad \bar{\eta}_q^{\cI\alpha}|0, p^+\rangle_{U,T} =0
\, ,  \ee
\be\label{fermivac}
{\theta}_R|0, p^+\rangle_{U}=0\, ,      \qquad \bar{\theta}_L|0, p^+\rangle_{U}=0\, .
\ee
As it was outlined in \cite{MT}, the above choice for the fermionic zero-mode vacuum is the one which reflects the $SO(4)\times SO(4)$ symmetry of the pp-wave background.

\subsubsection*{The Hamiltonian in the untwisted sector}
In the above basis  the \lc Hamiltonian in the untwisted sector reads (we restore here the $\alpha'$ dependence):
\be\label{zer} H_{U} = H_{0,U} +  {1 \ov \a'
p^+ } \sum_{\cI=1,2} \sum_{n=1}^\infty {\textstyle{ \sqrt{ {n^2}
+ (\a' p^+  \f)^2 }}} \    (a_n^{\cI I} \bar{a}_n^{\cI I}+
\eta_n^\cI\bar{\gamma}^-\bar{\eta}_n^\cI ) \ .
 \ee
The untwisted zero-mode Hamiltonian $H_{0,U}$ gets contributions from the 
normal ordering of fermionic and bosonic zero-modes. The 
normal ordering of the non-zero modes cancels between bosons and fermions, 
which appear in equal number. The general expression of $H_{0,U}$ is:
\be
H_{0,U}= \f ( a_0^I\bar{a}_0^I +
\theta_L^{}\bar{\gamma}^-\bar{\theta}_L^{} - \theta_R^{}\bar{\gamma}^-\bar{\theta}_R^{} + e_0).
\ee
The choice (\ref{fermivac}) for the vacuum amounts to setting
$e_0=0$ \cite{MT}.

In summary, the Hamiltonian in the type 0B untwisted sector is exactly the same as in the type IIB case. Differences will appear in the space 
of physical states.
\subsubsection*{The Hamiltonian in the twisted sector}
In the twisted sector we get the following expression for the \lc Hamiltonian:
\be\label{Tzer} H_{T} =  H_{0,T} + {1 \ov \a'
p^+ } \sum_{\cI=1,2} \left[\sum_{n=1}^\infty {\textstyle{ \sqrt{ {n^2}
+ (\a' p^+  \f)^2 }}} \    a_n^{\cI I} \bar{a}_n^{\cI I}+
\sum_{r=1/2}^\infty {\textstyle{ \sqrt{ {r^2}
+ (\a' p^+  \f)^2 }}} \ \eta_r^\cI\bar{\gamma}^-\bar{\eta}_r^\cI \right] \ .
 \ee
The zero-point Hamiltonian $H_{0,T}$ now gets a non-trivial contribution from the normal ordering of the fermionic and bosonic non-zero modes, bacause the fermions here obey antiperiodic boundary conditions. Explicitely we have $H_{0,T}=\f (a_0^I\bar{a}_0^I) + E_{0,T}$, with:
\be
E_{0,T} = 4\f  + {8 \ov\a'
p^+ } \left[\sum_{n=1}^\infty {\textstyle{ \sqrt{ {n^2}
+ (\a' p^+  \f)^2 }}}- \sum_{r=1/2}^\infty {\textstyle{ \sqrt{ {r^2}
+ (\a' p^+  \f)^2 }}}\right],
\ee
where the first addendum is the normal ordering constant of the $8$ bosonic zero modes.

In order to evaluate the difference of the diverging series in the previous expression, we make use of the {\it Epstein function} (the explicit representation we consider here is one of the possible and is taken from \cite{Albu}):
\begin{eqnarray}
F[z,s,m^2]&=& \sum_{n=1}^{\infty}\left[(n+s)^2 + m^2\right]^{-z}= {1\over2}[(s+1)^2 +m^2]^{-z} +
\int_1^{\infty}dx[(x+s)^2+m^2]^{-z} + \nonumber \\
&+& i\int_0^{\infty}dt\left[{ [(1+it+s)^2+m^2]^{-z}-
[(1-it+s)^2+m^2]^{-z}}\over{e^{2\pi t}-1}\right].
\end{eqnarray}
This has a simple pole when $z=1/2, -1/2, -3/2, ...$. We
are interested in $z=-1/2$. The pole cancels in $E_{0,T}$
due to the opposite contribution of bosons and fermions. The
result is  (let us note that $s=0$ for the bosonic series,
and $s=-1/2$ in the fermionic case): \begin{eqnarray} E_{0,T}&=&
4\f
-{1\over{\alpha'p^+}}\left[\sqrt{1+4(f\alpha'p^+)^2}\right. \nonumber \\
&&\left. +
4(f\alpha'p^+)^2\log\left[{2+2\sqrt{1+(f\alpha'p^+)^2}}\over{1+
\sqrt{1+4(f\alpha'p^+)^2}}\right]-8I(f\alpha'p^+)\right],\label{zert}
\end{eqnarray} where: \be I(f\alpha'p^+)= i\int_0^{\infty}dt\left[{f(0;1+it)-
f(0;1-it)- f(-1/2;1+it)+f(-1/2;1-it)}\over{e^{2\pi t}-1}\right],
\label{in} \ee and: \be f(s;x)=\sqrt{(x+s)^2+(f\alpha'p^+)^2}. \ee
In the flat space limit $f=0$, we recover the standard type 0B
zero-point light-cone Hamiltonian: \be H_{0,T}(f=0)=
-{1\over{\alpha'p^+}}, \ee which signals the presence of a
tachyonic field of mass squared $m^2= -2/\alpha'$ corresponding to
the Fock vacuum of the twisted sector.

For $f\neq 0$, $E_{0,T}$ is a non-analytic function of
$g^{-1/2}_{eff}\sim(\alpha'p^+f)$. For $g_{eff}=0$ it is zero, and
this value does not receive any (``perturbative'') correction in
powers of $g_{eff}$: order by order the energy
is zero. The twisted zero-mode energy is thus non-perturbative in $g_{eff}$ \footnote{
  The nature of the ``non perturbative'' corrections is
 clearer in the form of the analytic continuation of $F[z,s,m^2]$
 that can be found, for example, in \cite{elizalde}. In that paper
 the corrections take the form of Bessel functions, with an
 exponential vanishing behavior.}. 
$E_{0,T}$ is always negative, approaching zero for $g_{eff}\rightarrow0$ and $4f-{1\over{\alpha'p^+}}$ for $g_{eff}\rightarrow\infty$.

\section{The spectrum of physical states}
Generic Fock space vectors in both sectors are built up in terms
of  products of  creation operators acting on the vacuum. The
subspace of physical states  is obtained by imposing the
constraint (\ref{zeromoml}), which in terms of the oscillators
reads \cite{MT}: \be
 N^1|\Phi_{phys}\rangle
=N^2|\Phi_{phys}\rangle \  , \ \ \ \ \ \ \ \ \
 N^\cI = \sum_{n=1}^\infty k_n a_n^{\cI I}\bar{a}_n^{\cI
I} + \sum_{q}k_q \eta_q^\cI\bar{\gamma}^-\bar{\eta}_q^\cI \ . \ee
In the type 0B model, however, this is not the only constraint to impose.
In fact,  the spectrum of physical states is obtained by projecting out all the states which are not invariant under the orbifold projection $\theta\rightarrow -\theta$. So, just as in the flat case, {\it we will not have spacetime fermions in the spectrum}: all the states obtained from the vacuum by acting with an odd number of fermionic operators are projected out.
\\

In the {\bf untwisted} sector the generic state built only in
terms of fermionic zero modes is: \be
(\bar{\theta}_R)^{n_R}(\theta_L)^{n_L}|0,p^+\rangle_U , \quad   n_L,
n_R= 0, 1, 2, 3,4, \label{gen} \ee where the limited range of
values of $n_R, n_L$ is due to the relations
$(\bar{\theta}_R)^5=(\theta_L)^5= 0$. Also, the type 0 projection
requires $n_L+n_R$ to be even.
Acting with the Hamiltonian operator on the states (\ref{gen}) we
find: \be H_U(\bar{\theta}_R)^{n_R}(\theta_L)^{n_L}|0,p^+\rangle_U
=\f( n_R + n_L)(\bar{\theta}_R)^{n_R}(\theta_L)^{n_L}|0,p^+\rangle_U
. \ee

Thus the lowest energy state is just the Fock vacuum which has
zero light-cone energy. This state corresponds to a purely gravity
mode which is obtained from a linear combination of the trace of
the graviton and the four-form potential \cite{MT}. The other
states, with energies $2f,4f,6f,8f$ (which are all zero in the
flat limit), complete the untwisted gravity sector. 
The action of the bosonic zero modes gives for each state a tower of 
states with higher energies.  
Unlike the zero-mode sector, all the higher oscillators produce
states whose light-cone energy is a function of
$g^{-1/2}_{eff}\sim (f\alpha'p^+)$. Their energy becomes infinite in the
gravity limit
$g_{eff}\rightarrow\infty$, and they decouple from the gravity modes, 
as appropriate for stringy modes in a weakly curved background. 
On the other hand, zero modes and excited oscillator states are almost
degenerate in the highly curved background limit $g_{eff}\rightarrow 0$. This
is the regime where comparison with a perturbative gauge theory can be done.

The spectrum of physical states is just the bosonic part
of the corresponding type IIB spectrum.
\\

In the {\bf twisted sector}, as we have anticipated, the Fock
vacuum energy $E_{0,T}$ (\ref{zert}) depends non-trivially on $g_{eff}$ and is always negative. In
the gravity limit $g_{eff}\rightarrow\infty$ \be
E_{0,T}{|0,p^+\rangle}_T \approx (4f -
{1\over\alpha'p^+}){|0,p^+\rangle}_T \label{bret}\ee 
and we can interpret the two terms in the
previous formula as the effect of the curved spacetime ($4f$) and
the contribution of the ``mass'' squared of a tachyonic scalar field 
($-2/\alpha'$). 
In the Appendix formula (\ref{bret}) is derived in the gravity
approximation. In type 0B, the negative mass squared of the tachyon
in the gravity limit $g_{eff}\rightarrow 0$ is so large to 
overwhelm the positive
contribution of the curvature, and we are left with an 
infinitely negative twisted ground state energy.
When $f=0$ this state corresponds to the usual type 0B tachyon. 
As in the untwisted sector, the action of the bosonic zero modes gives a 
tower for each state in the spectrum.

The first excited states are obtained by acting on the vacuum with 
$\eta^1_{1/2}\eta^2_{1/2}$. The corresponding value of the light-cone 
Hamiltonian depends non-trivially on $g_{eff}$, but now it is always positive.
In the gravity limit $g_{eff}\rightarrow\infty$ their energy goes to the 
finite value $4f$. 
When $f=0$ this is zero and these states correspond to 
massless fields of the twisted R--R sector. 
Note that the untwisted sector R--R 
fields have energies that are independent of $g_{eff}$ and are different
from each
other. On the other hand, the energies of the twisted R-R forms are equal
and vary with $g_{eff}$. 
The two sets of R-R fields appear to be quite different, 
so it would be interesting to study if there exist combinations 
coupling to electric and magnetic branes as in the flat space case.
All other excited string states decouple in the gravity limit.

In the opposite limit $g_{eff}\rightarrow0$ the energy of the tachyon is zero,
and the energy of the R--R fields is $2f$. Just as in the untwisted sector, 
all string states are almost degenerate. 
In this regime,
we can expect to be able to compare the string theory 
with the field theory results.

\section{Field theory interpretation}
The dual to the type 0B String on $AdS_{5} \times S^{5}$ with self
dual R--R five form flux is conjectured to be a large N conformal
non supersymmetric gauge theory
 with group SU(N) $\times$ SU(N) that, apart from the gauge bosons, consists of six scalars
 $\phi^i$, $\tilde{\phi}^i$, $i=1,...,6$, (the tilde distinguishes the two gauge groups) in the adjoint representation
  of each of the two groups (that have the same coupling
 constant), four bifundamental Weyl spinors $\chi^{\alpha}$ in the $(\textrm{N},\bar{\textrm{N}})$
 and four, $\psi^{\alpha}$, in the $(\bar{\textrm{N}},\textrm{N})$ \cite{KT}.\\
This theory can be obtained from the $\cal{N}$$= 4$ SU(2N) SYM
by a $Z_2$ projection, just as type 0B can be obtained from an orbifold of
type IIB. If we arrange the $\cal{N}$$= 4$ fields in 2N $\times$
2N matrices, the projection is made with $(-1)^F$, where $F$ is the
fermion number, together with a conjugation by
$\cal{I}$$=diag(I,-I)$, $I$ being the N $\times$ N identity
matrix \cite{nek}. As was shown in \cite{nek} this construction implies that
the orbifold group lives in the center of SU(4).
This leaves the diagonal N $\times$ N blocks for the bosons
and the off-diagonal ones for the fermions. The operators of the
theory can be organized in untwisted and twisted ones,
depending on the behavior under the projection: the formers are
even under exchange of the two gauge groups, the latter are odd.
They can be written as traces of products of the 2N $\times$ 2N matrices
representing the ${\cal N}= 4$ fields, inserting $\cal{I}$ when dealing
with twisted operators. 
For example, the  ${\cal N}= 4$
operator $\chi Z \chi Z$, 
gives rise in the twisted sector to operators of the form:
\begin{equation}
tr \left[\left(
\begin{array}{cc} 0 & \chi
\\ \psi & 0 \end{array}
\right) \left( \begin{array}{cc} Z & 0 \\ 0 & \tilde{Z}
\end{array} \right)\left(\begin{array}{cc} 0 & \chi\\
\psi & 0 \end{array} \right)\left( \begin{array}{cc}  Z & 0 \\ 0 &
\tilde{Z}
\end{array} \right)\cal{I}\right]=\chi
\tilde{Z} \psi Z-\psi Z \chi \tilde{Z},\label{matrix}
\end{equation}
where we omitted the trace over the gauge indexes. Our notations follow 
\cite{MBN}, so, for example, $Z={1\over
\sqrt{2}}(\phi^5+i\phi^6$).\\

The untwisted sector inherits the
non-renormalization properties of the parent $\cal{N}$$=4$ theory.
In fact, the correlators of the untwisted sector operators in the planar limit
 are exactly the same as the corresponding correlators in
 $\cal{N}$$=4$ SYM \cite{vafa,nek}. The identification of operators with 
string states in the untwisted sector therefore is 
the same as that for the bosonic part of the type IIB model.

The twisted sector is not protected by any $\cal{N}$ $ =4$ heritage.
For large 't Hooft coupling constant $\lambda$, the $AdS$ spectrum
contains an operator with complex
dimension, making the whole theory unphysical \cite{K}. This
operator is precisely the one whose source is the bulk tachyon.
It has also been argued in \cite{AS} that the theory is 
unstable also for small $\lambda$ due to a scalar
potential destabilizing the vacuum. 
The pp-wave limit corresponds to large $\lambda$ $and$ large $J$
values. In this regime, as anticipated in the Introduction, we don't expect 
any operator with
complex dimension. The reason is that only high harmonics (with
charge $J$) of the tachyon survive the limit. These highly excited states
have an effective mass squared (measured in units of the $AdS$ radius) 
of order $-2\sqrt{\lambda}  + 16 + J(J+4)$, 
where the first term is the flat space value for the tachyon mass, 
the second one comes from the coupling with the $AdS$ five 
form\footnote{Notice
that in the pp-wave background the five 
form does not give any contribution 
to the mass of the tachyon.} and the third one comes from the angular momentum
 on the five-sphere.
Recalling that $J\sim \sqrt{\lambda}$, the masses are 
expected to satisfy
the Breitenlohner-Freedman bound on $AdS$. 
If we (naively) extrapolate the
mass/dimension formula of the $AdS/CFT$ dictionary to the pp-wave
regime, we are left with positive
 dimension operators.
In fact we do not encounter any instability in $\Delta$, 
which is positive and of order $\sqrt{\lambda}$.
We find instead  that $\Delta -J$ is unbounded from below. 
This phenomenon presumably descends from the instabilities of
the original theory, but we have no reliable analysis at hand.

\subsection{Untwisted sector}
 The string ground
 state has $H_U = 0$. The only operator in the untwisted sector with $\Delta - J=0$ is $Tr[Z^J] + Tr[\tilde{Z}^J]$. In the
  planar limit its dimension is expected
 to be protected as for the corresponding BPS $\cal{N}$$=4$ operator. We identify it with the string ground state:
\begin{equation}
|0,p^+\rangle _U \quad \longleftrightarrow \quad {1\over
\sqrt{2JN^J}}(Tr[Z^J] + Tr[\tilde{Z}^J]). \label{groundu}
\end{equation}
The identification of the states constructed acting on the ground
state with the zero modes and with the oscillators is the same
as that in the type IIB
theory. 
The action of the zero modes on the vacuum corresponds to operators whose dimension is also protected. As a general rule, in both the untwisted and twisted sectors, the action of the bosonic
 oscillators $a^{{\cal{I}}I}$ for $I=5,...,8$ corresponds to
 insertions in the correlators of the fields $\phi^{I-4}$, while
 for $I=1,...,4$ to insertions of $D_{I}Z$. The insertions of fields with $\Delta - J > 1$ give rise
 to operators that decouple in the pp-wave limit.
As an example of zero mode insertion we have:
\begin{equation}
a_0^{7}|0,p^+\rangle _U \quad \longleftrightarrow \quad {1\over
\sqrt{2N^{J+1}}}(Tr[\phi^3 Z^J] + Tr[\tilde{\phi^3}\tilde{Z}^J]).
\label{zerou}
\end{equation}
 In the same way, as an example of the first excited string states, one
 has:
\begin{equation}
a_n^{1,8}a_{n}^{2,7}|0,p^+\rangle _U \quad \longleftrightarrow
\quad {1\over \sqrt{2JN^{J+2}}}(\sum_{l=0}^J Tr[\phi^4 Z^l \phi^3
Z^{J-l}]e^{2\pi i n l \over J}+\sum_{l=0}^J Tr[\tilde{\phi}^4
\tilde{Z}^l \tilde{\phi}^3 \tilde{Z}^{J-l}]e^{2\pi i n l \over
J}). \label{oscu}
\end{equation}
As in the $\cal{N}$$=4$ case, the classical dimension of the stringy states is
perturbatively corrected. The perturbative series in $g_{eff}$ is
re-summed by the
 square root in (\ref{zer}).\\

The action of the fermionic oscillators on the vacuum,
corresponding to the insertion of the bifundamental spinors,
involves combinations of factors of the form $Tr[\chi \tilde{Z}^l
\psi Z^{J-l}]$ (the projection forces always the presence of an
even number of spinors, giving only bosonic operators). For
example, the vacuum state acted on by two fermionic zero-modes
corresponds to the phase-unweighted operator\footnote{It has protected dimension. This is not the case for the ``minus-sign'' counterpart, which will receive perturbative corrections in $\lambda$ and will decouple in the pp-wave limit.}:
\begin{equation}
\bar{\theta}_R^{a}\theta_{L}^{b}|0,p^+\rangle _U \quad
\longleftrightarrow \quad {1\over \sqrt{2JN^{J+2}}}(\sum_{l=0}^J
Tr[\chi^a \tilde{Z}^l \psi^b Z^{J-l}]+\sum_{l=0}^J Tr[\psi^a Z^l
\chi^b \tilde{Z}^{J-l}]), \label{ferzerou}
\end{equation}
where $a,b=1,...,8$, on the operator side count only the
$J={1\over 2}$ components of the spinors.\\
The excited string states correspond to:
\begin{equation}
\eta_n^{1,a}\eta_{n}^{2,b}|0,p^+\rangle _U \quad
\longleftrightarrow \quad {1\over \sqrt{2JN^{J+2}}}(\sum_{l=0}^J
Tr[\chi^a \tilde{Z}^l \psi^b Z^{J-l}]e^{2\pi i n l \over
J}+\sum_{l=0}^J Tr[\psi^a Z^l \chi^b \tilde{Z}^{J-l}]e^{2\pi i n l
\over J}). \label{feroscu}
\end{equation}

The first correction in $g_{eff}$ to
the dimension of the operators (\ref{oscu}) and (\ref{feroscu})
has the correct behavior to
reproduce the first term in the expansion of the square root in 
(\ref{zer}). All relevant contributions to the anomalous
dimensions come from contracting each term in (\ref{oscu}) and 
(\ref{feroscu}) with itself. Possible off-diagonal contributions 
-- for example, the contraction of the $l$ terms in the first sum 
in (\ref{feroscu}) with  the $J-l$ term in the second sum -- 
are suppressed by a $1/J$ factor, due to the phase factors. 
The calculation is then identical
to the one in \cite{MBN}. Graphs with fermions contains two
Yukawa interactions.

\subsection{Twisted sector}
The energy of the string ground state, given by formula (\ref{zert}),
is a function of $(f \alpha' p^+ )^2 \sim {1\over g_{eff}}$, as we
have seen. For $g_{eff}=0$ it is zero. The function $E_{0,T}$ is
not analytic in $g_{eff}$, and there are no power corrections 
in $g_{eff}$ around $g_{eff}=0$. This
means that the dual operator has no perturbative corrections (in
$g_{eff}$) to its classical dimension. $\Delta -J$ is therefore
zero. The operator is:
\begin{equation}
|0,p^+\rangle _T \quad \longleftrightarrow \quad {1\over
\sqrt{2JN^J}}(Tr[Z^J] - Tr[\tilde{Z}^J]). \label{groundt}
\end{equation}
Note that in $AdS$ the operator dual to the tachyon is identified with
$Tr(F_1^2)-Tr(F_2^2)+...$.  
In a sense, the $AdS$ tachyon is the twisted
sector counterpart of the operator dual
to the dilaton $e^{\phi}\rightarrow Tr(F_1^2)+Tr(F_2^2)+...$. 
In the pp-wave background, instead, the operator dual to the tachyon contains only the $Z$ scalars\footnote{The pp-wave dilaton is identified
with the untwisted ground state acted upon by four fermionic
zero modes \cite{MT}. In $AdS$ a dilaton harmonic with charge $J$ is
dual to an operator containing, among others, terms
of the form $Tr(F^2Z^{J-2}+\chi^4Z^{J-4}+...)$. In the large $J$ limit,
the leading contributions to correlation functions come from the terms
with arbitrary insertion of
four fermions in the string of $Z's$, consistently with the string picture.}. This is consistent with the fact that the zero-mode of 
the $AdS$ tachyon
couples to $F^2$, while the higher
harmonics couple to scalar fields \cite{garousi,AS}. This is also
manifest in the form of the Born-Infeld Lagrangian in Einstein frame:
\begin{equation}
\int d^4x V(T) \sqrt{|g+e^{-\phi/2}F|}
\end{equation}
when the tachyon field is Taylor expanded, $T=\sum T_n \phi_{\{i_1}...\phi_{i_n\}}$.

Reconstructing the function (\ref{zert}) in field theory is not an
easy task. First of all, as we have seen, we need to consider non
perturbative $g_{eff}$ corrections. Moreover we should work in the
strongly coupled $\lambda$ regime. Note that, being
(\ref{zert}) non analytic, we don't expect to be able to extrapolate
the perturbative $\lambda$ calculations to the strong regime; of
course, we can't rely on supersymmetry to protect our results.
As a check of the validity of the identification, we
note that to loop order $J$ the operator has no perturbative corrections. 
Any difference
with the untwisted operator (\ref{groundu}), which is protected, 
resides in interactions of the form
$Tr[Z^J]Tr[\tilde{Z}^J]$, 
that give the first contributions at order $\lambda^J$. 
Contribution of this order are problematic 
in the BMN approach and are usually neglected \cite{gross}. 
We interpret our result as an indication that these $\lambda^J$ corrections 
disappear in the large 't Hooft coupling regime.\\

Operators in the twisted sector contain an extra minus-sign with respect
to the untwisted sector ones, due to the
insertion of the matrix $\cal{I}$. 
The duals to states with bosonic oscillators are the minus-sign
duplicates of the ones in (\ref{zerou}), (\ref{oscu}).
The fermionic oscillators are instead
half-integer modded, so in the identification there are different
phases with respect to the untwisted sector. The first excited
string states ($r=1/2$) correspond to the R--R forms and are made
with two fermions acting on the vacuum. In field theory we have, for
generic half-integer $r$:
\begin{equation}
\eta_r^{1,a}\eta_{r}^{2,b}|0,p^+\rangle _T \quad
\longleftrightarrow \quad {1\over \sqrt{2JN^{J+2}}}(\sum_{l=0}^J
Tr[\chi^a \tilde{Z}^l \psi^b Z^{J-l}]e^{2\pi i r l \over
J}-\sum_{l=0}^J Tr[\psi^a Z^l \chi^b \tilde{Z}^{J-l}]e^{2\pi i r l
\over J}). \label{ferosct}
\end{equation}
In the energy formula (\ref{Tzer}) for these states we have two separate contributions from the zero
point energy (the energy of the tachyonic ground state) 
and the square root. The zero point energy gives no perturbative
contribution, as we have seen.
The square root can be easily reproduced at first order in field theory,
the calculation being exactly the same as that in the untwisted sector.
The only difference is the half-integer phase factor.
For example, for the first excited state $r=1/2$, we have
a factor of ${1\over 2}$ in the argument of
the exponentials, whose square gives the ${1\over 4}$ difference
in the first nontrivial term in the square root.

Finally, we notice that, for operators with bosonic insertions,
the relative sign in (\ref{oscu}) discriminates between untwisted
and twisted sector. For operators with fermionic insertions,
due to the difference in modding of the phase factors, 
we can write two more operators
by reversing signs in (\ref{feroscu})
and (\ref{ferosct}). These operators must have anomalous
dimensions that diverge for large 't Hooft coupling, since there are
no string states corresponding to them. Notice that an operator
is almost-BPS only if all contributions in $\lambda$ to the anomalous
dimension cancel, even those that are usually neglected in the
dilute-gas approximation because suppressed by powers of $J$.
In our case, already at first
order in $\lambda$, there are ${\lambda\over J}\sim \sqrt{\lambda}$ 
contributions to the anomalous dimensions that are canceled only
with a specific choice of signs.
To check this, it is convenient to write the generic
operator with two fermionic impurity in $\cal{N}$$=4$ notations as:
\begin{equation}
O_{k,q}={1\over\sqrt{2JN^{J+2}}} \sum_{l=0}^JTr[\chi^1Z^l\chi^2Z^{J-l}{\cal I}^k]e^{{2\pi iql\over J}},
\end{equation} 
where $\chi^i$ and $Z$ are 2N$\times$ 2N matrices, $k=0,1$.
$\cal{I}$
commutes with all $Z$'s and anticommutes with $\chi^i$.
Most of the leading contribution in 
$\langle O_{k,q}\bar O_{k,q} \rangle$ come from 
contracting the $l$ term in the sum with itself. However, in the BPS
case $q=k=0$, a ${\lambda\over J}$ contribution, coming from the exchange
graphs for the two impurities, cancels when summing the $l=0/l=0$, $l=J/l=J$ and $l=0/l=J$ contractions. For generic $q$ and $k$, the  $l=0/l=J$ 
contraction carries an extra $(-1)^ke^{2\pi iq}$ factor, where
the $(-1)^k$ is due to the fact that $\cal{I}$ anticommutes with the
impurities. We thus obtain an exact cancellation of the ${\lambda\over J}$
terms only for $(-1)^ke^{2\pi iq}=1$. This correlates the integer
or half-integer modding of the phase with the insertion of $\cal{I}$.


\vskip .2in \noindent \textbf{Acknowledgments}\vskip .1in
\noindent We would like to thank Sergio Cacciatori, Claudio Destri 
and Arkady Tseytlin for useful discussions and comments.
F.B. is partially
supported by INFN and the
European Commission TMR program HPRN-CT-2000-00122. L.G., A. L. C. and A. Z. are partially
supported by INFN and MURST under contract 2001-025492, and by the European Commission TMR
program HPRN-CT-2000-00131, wherein they are associated to the
University of Padova.

\setcounter{section}{0} \setcounter{subsection}{0}
\appendix{Appendix: Scalar fields stability bound in PP-wave}
In this appendix we reproduce 
the result for the mass of the tachyon in the gravity approximation\footnote{Here
and in the following we will use the fact that the non-trivial
components of the connection and  curvature  are ($g^{--}=
\f^2x_I^2$): $ \Gamma^\mun_{+I} = -\f^2x^I\delta_-^\mun \,,\qquad
\Gamma_{++}^\mun = \f^2x^I\delta_I^\mun\ ,  \ \ \ \ \
R_{I++J}=-\f^2\delta_{IJ}\,,\qquad R_{++}=8\f^2 \ $. In particular
the scalar curvature is zero in the background.}.

The ``massive'' scalar Lagrangian density in the background (\ref{pp}) reads ( ${\sqrt{-g}}= 1$):
\begin{equation}
{\cal {L}}= -{1\over2}\left[ 2\partial_+\phi\partial_-\phi + f^2x^2\partial_-\phi\partial_-\phi + \partial_I\phi\partial_I\phi + m^2\phi^2\right].
\label{lagra}
\end{equation}
When $f=0$ we recover the flat space Lagrangian in the front-form of dynamics, in which $x^+$ plays the role of time variable.


We define
the energy as the generator of $x^+$
translations. It is evident that $\xi_+ =
\xi_+^{\mu}\partial_{\mu}= \partial_+$ is a Killing vector for the
background $(\ref{pp})$, so that the following quantity ($T$ being
the stress tensor for gravity plus scalar fluctuations): \be
H=\int dx^- d^8x T^{+\mu}g_{\mu\nu}\xi^{\nu}_+ =  \int dx^- d^8x
T_{+-} \ee is (formally) conserved in $x^+$. The scalar field
contribution in $T_{+-}$ corresponds to the canonical Hamiltonian
density derived from (\ref{lagra}):
\begin{equation}
{\cal {H}}= \pi\partial_+\phi - {\cal {L}}={1\over2}\left[ f^2x^2\partial_-\phi\partial_-\phi + \partial_I\phi\partial_I\phi + m^2\phi^2 \right].
\end{equation}

Let us stress that, as noted in \cite{BF}, this does not
uniquely fix the energy functional. In fact we can define an
``improved'' stress tensor (which is also conserved), whose
relevant component reads:
\begin{equation}
T_{+-}^{im}= T_{+-} + \beta(\Box -\partial_+\partial_-)\phi^2.
\end{equation}
Here, referring to the notations in \cite{BF}, we have used the
fact that $D_+\partial_{-}\phi=\partial_+\partial_-\phi$, that
$R_{+-}=0$, and that, being $V(\phi)\approx m^2\phi^2$, in our case
the critical point is $\phi_0=0$ and so $V(\phi_0)=0$.
\subsection{Equation of motion and its solution}
The equation of motion following from (\ref{lagra}) is:
\begin{equation}
(\Box-m^2)\phi \equiv \left[2\partial_+ \partial_- +
f^2x^2\partial_- \partial_- + \partial_I\partial_I -
m^2\right]\phi = 0. \label{eqmot}
\end{equation}
This is of first order in $\partial_+$, and it is well now (see for example \cite{Heinzl}) that the associated Cauchy problem can be solved by imposing mixed initial-boundary conditions.
Here we are interested in those
solutions which go to zero when $x^I \rightarrow\infty$ (these are
required if we want our energy functionals to be actually
convergent and so effectively conserved in $x^+$). Then we choose (see
also \cite{Leigh}):
\begin{equation} \label{cho}
\phi\approx e^{-ip_+x^+}e^{ip_-x^-}e^{-cx^2},
\end{equation}
which is an acceptable solution\footnote{Let us outline that the very general
normalizable solution contains a product of Hermite polynomials depending on
the transverse coordinates. The choice (\ref{cho}) corresponds
to the lowest of such polynomials. We are interested 
in the string theory
vacuum, without any insertion of bosonic zero-mode
oscillators.} of (\ref{eqmot}) provided that:
\begin{equation}
c= {f|p_-|\over2};\quad   2p_+p_- - 16c - m^2=0.
\label{con}
\end{equation}
Thus the general form of the selected kind of solution reads:
\begin{equation}
\phi(x^+, x^-, x^I)= \int_{-\infty}^{\infty}dp \theta(p)\left[a(p)e^{-{i\over2p}(m^2+8fp)x^+}e^{ipx^-}+ a^*(p)e^{{i\over2p}(m^2+8fp)x^+}e^{-ipx^-}\right]e^{-{{fp}\over2}x^2}.
\label{sol}
\end{equation}
Here we have relabelled $p_-=p$ and imposed the reality condition on $\phi$.

From the relations (\ref{con}) we can see that in the gravity
limit $\alpha'p^+f\rightarrow 0$ the vacuum state in the twisted
sector of our type 0B model (whose light cone energy is $p_+\approx 4f-
(1/\alpha'p^+$)) can be viewed as a scalar fluctuation of mass $m^2= -
(2/\alpha')$.

\subsection{The energy functional}
Let us now evaluate the Hamiltonian on the solution (\ref{sol}).
After integration in $x^-$ we find:
\begin{equation}
H = \int d^8x\int dp\theta(p)a(p)a^*(p)[ 2f^2p^2x^2 +
m^2]e^{-fpx^2}.
\end{equation}
The integration over $x^I$ is easily performed in spherical coordinates
using the fact that (for $a>0$):
\begin{equation}
\int_0^{\infty}dy y^n e^{-ay}= {n!\over a^{(n+1)}}.
\end{equation}
We see that: 
\begin{equation}
H= 3\Omega\int_0^{\infty}{dp\over(fp)^4}[8pf + m^2]a(p)a^*(p).
\end{equation}
($\Omega>0$ results from the integration over the transverse angular 
variables).

Each mode of momentum $p_- (= p^+)$ gives a positive contribution to the energy, provided that:
\begin{equation}
m^2 > -8(fp^+).
\label{bound1}
\end{equation}
In particular we see that the state corresponding to the twisted sector 
vacuum has, in the gravity limit, a mass which violates this bound, 
giving a negative light-cone energy.

Let us now turn to the ``improved'' energy functional:
\begin{equation}
H_{im} = H + \beta\int dx^-d^8x(\Box -\partial_+\partial_-)\phi^2.
\end{equation}
By using the equation of motion we find that:
\begin{eqnarray}
H_{im}= (4\beta +1)H + 2\beta\int dx^-d^8x(\partial_+\phi\partial_-\phi- \phi\partial_+\partial_-\phi).
\end{eqnarray}
On the solution (\ref{sol}) we have:
\begin{equation}
H_{im}= H.
\end{equation}
We see that the improved energy does not give any other bound 
on the mass parameter.

\end{document}